\documentclass[reprint,
twocolumn,
superscriptaddress,
graphicx,
floatfix,
bibnotes,
 amsmath,
 amssymb,
aps,
prl,
]{revtex4-2}

\usepackage{chemformula} 
\usepackage[T1]{fontenc} 
\usepackage{amssymb}
\usepackage{braket}
\usepackage{float}

\usepackage{ulem}

\usepackage{color}

\usepackage{graphicx}
\usepackage{dcolumn}
\usepackage{bm}
\usepackage{booktabs}
\usepackage{multirow}

\begin{document}

\preprint{}

\title{Role of interlayer shear phonons on lattice symmetry switching in the transition metal dichalcogenide WTe$_{2}$}

\author{Mizuki Akei}
\email{s2420263@u.tsukuba.ac.jp}
\affiliation{Department of Applied Physics, Graduate school of Pure and Applied Sciences, University of Tsukuba, 1-1-1 Tennodai, Tsukuba 305-8573, Japan}
\author{Takumi Fukuda}
\email{takumi.fukuda@oist.jp}
\affiliation{Femtosecond Spectroscopy Unit, Okinawa Institute of Science and Technology Graduate University, 1919-1 Tancha, Onna, Okinawa, Japan}
\author{Yu Mizukoshi}
\affiliation{Department of Applied Physics, Graduate school of Pure and Applied Sciences, University of Tsukuba, 1-1-1 Tennodai, Tsukuba 305-8573, Japan}
\author{Kazuhiro Kikuchi}
\affiliation{Department of Applied Physics, Graduate school of Pure and Applied Sciences, University of Tsukuba, 1-1-1 Tennodai, Tsukuba 305-8573, Japan}
\author{Muneaki Hase}
\email{mhase@bk.tsukuba.ac.jp}
\affiliation{Department of Applied Physics, Graduate school of Pure and Applied Sciences, University of Tsukuba, 1-1-1 Tennodai, Tsukuba 305-8573, Japan}

\date{\today}

\begin{abstract}
Coherent phonon control using ultrashort pulse trains is the key to realizing structural phase transitions in solids by non-thermal pathways. By combining double-pulse excitation and time-resolved second harmonic generation techniques under high-density electronic excitation in a 2D layered material, WTe$_{2}$, we demonstrate that the lattice symmetry switching from the Weyl semimetallic $T_{d}$ to the semimetallic 1$T^{\prime}$ phases is independent of the amplitude of the coherent interlayer shear phonons after the arrival of the second pump pulse. This finding provides new insights into the mechanisms for symmetry switching that electronic excitation-driven shear sliding plays a dominant role.
\end{abstract}

\maketitle


\newpage

Photons excite multiple degrees of freedom in matter and induce cooperative phase transitions between electronic and lattice systems via non-thermal pathways. This phenomenon, known as photo-induced phase transition (PIPT) \cite{KOSHIHARA20221}, has been a prominent theme in ultrafast spectroscopy applied to topics ranging from photochemistry to solid-state physics for several decades. 
In classical (incoherent) PIPT, a new phase emerges due to energy relaxation of electronic excited states \cite{KoshiharaPRB1990}, while in quantum (coherent) PIPT, a transition to a new phase occurs due to coherent control of the electronic or lattice system \cite{Iwai2006PRL,Horstmann2020Nature}. Variable time-resolved measurements have made it possible to realize and observe PIPT in liquid crystals \cite{hada2019ultrafast}, phase-change materials \cite{Makino2011Opex}, and correlated materials \cite{Perfetti2006PRL, Fausti2011Science}. Furthermore, control of ultrafast phase transition based on coherent PIPT is crucial not only for understanding the origin of phase transitions but also for applications such as terahertz frequency-driven optoelectronic memory devices \cite{Xiao2020Berry}. 
One promising approach for coherent PIPT is optical pulse-train excitation by tuning the time interval between pump pulses \cite{weiner1990femtosecond, hase1996APL, JHKim2009PRL}. This excitation scheme allows for the exploration of far non-equilibrium states and coherent control of electronic and structural phases, which is not possible with single-pulse excitation. For example, sub-picosecond electron interference and scattering dynamics \cite{hu2012delayed,Kimata2020PRBCoherent,iwasaki2023electronic}, photo-induced hidden states \cite{yusupov2010NatPhys,Zong2021PRL,mizukoshi2023APL, maklar2023SciAdvHidden}, and optical control of structural phases \cite{hase2015femtosecond,Horstmann2020Nature} have been investigated.

Recently, a feasible system for optical control of coherent PIPT involves ultrafast lattice symmetry switching between the semimetallic 1$T^{\prime}$ (centrosymmetric) and Weyl semimetallic $T_{d}$ (non-centrosymmetric) phases of layered transition-metal dichalcogenides (TMDs) such as $\mathrm{WTe_2}$ and $\mathrm{MoTe_2}$ \cite{sie2019ultrafast, Zhang2019PRX, fukuda2020ultrafast, cheng2022persistent}. Notably, driving a coherent interlayer shear phonon displacement provides a means to control the lattice symmetry and allow periodic alternation of adjacent structural phases at sub-THz frequencies. To realize coherent PIPT for optical control, it is essential to elucidate the dynamics of incoherent and coherent PIPT in TMDs, focusing on the effect of interlayer shear displacement. 
Additionally, under high-density electronic excitation conditions exceeding several $\mathrm{mJ/cm^2}$ by a single pulse, the intended phase transition may be suppressed by sample damage or saturable absorption \cite{SelenePRB2018, fukuda2022photo, fukuda2023UED2HMoTe2}. 
These issues can arise not only in controlling lattice symmetry but also in realizing optical control of other coherent PIPTs, such as spin states \cite{Okimoto2017PRA}, and must be addressed.

In this paper, we address these issues through experimental studies on lattice symmetry change under high-density electronic excitation using time-resolved second harmonic generation (TR-SHG), a sensitive probe for tracking symmetry change \cite{sie2019ultrafast,Zhang2019PRX,aoki2022excitation,hu2023strong,bykov2015coherent}.
By combining double-pulse excitation and TR-SHG methods in a two-dimensional layered material system, WTe$_{2}$, we show that the symmetric switch from the broken-inversion Weyl semimetal $T_{d}$ phase to inversion-symmetric semimetal 1$T^{\prime}$ phase is independent of the amplitude of the coherent shear phonons, which has been believed to play a central role \cite{sie2019ultrafast}.

The sample used in this study was a bulk single crystal of (001) surface $T_{d}$-WTe$_{2}$ (from HQ Graphene) with a thickness of $\sim$100 $\mu$m. The sample surface was cleaved with an adhesive tape prior to measurements. 
SHG arises from a non-zero second-order susceptibility in materials without inversion symmetry \cite{shen2003principles,Chang1997PRL,Glinka2011PRB}. In WTe$_{2}$, the $T_{d}$ phase has no inversion symmetry, whereas the 1$T^{\prime}$ phase does. Consequently, the $T_{d}$ phase exhibits SHG, while the 1$T^{\prime}$ does not \cite{sie2019ultrafast}. 
Note that when thin-layered samples are used, the contribution from the surface and/or interfaces can be prominent \cite{ZHAO2021105752}. 
However, the surface contribution would appear as a background with little variation with respect to the lattice symmetry switching, since the SHG responses are allowed in both the surface and bulk regions \cite{Zhang2019PRX,shen2003principles}.

\begin{figure}
   \begin{center}
	\includegraphics[width=8.4cm]{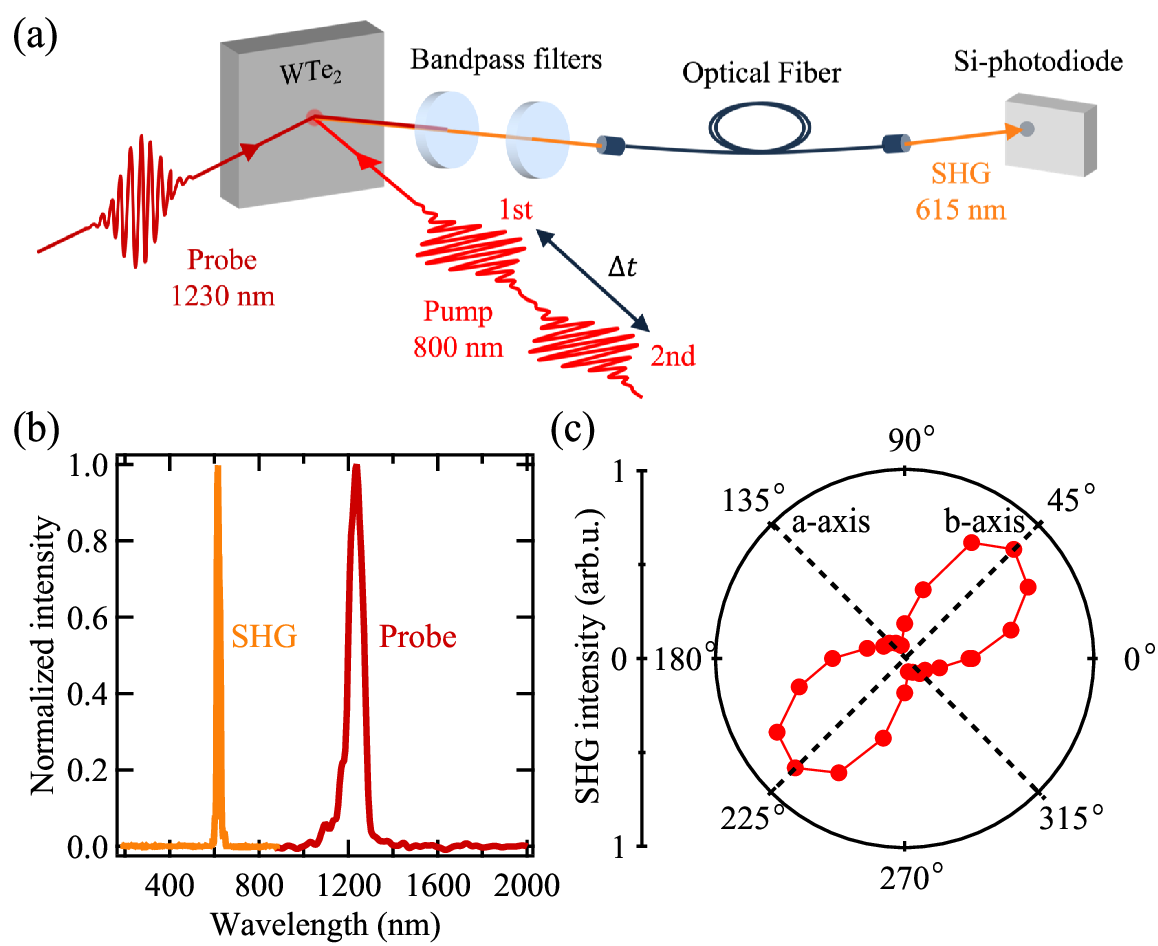}
	\caption{
	(a) Schematic of the time-resolved SHG measurements with a reflection geometry. The time interval between the 1st and 2nd pump pulse is expressed as $\Delta t$.
	(b) Measured spectra for the incoming probe light (1230 nm) and induced SHG (615 nm) from $T_{d}$-WTe$_{2}$ at the probe fluence of 3.2 mJ/cm$^{2}$.
	(c) Polarization angle dependence of static SHG intensity. }
	\label{Fig1.eps}
	  \end{center}
\end{figure}

The optical setup for the TR-SHG measurements with a reflective geometry is shown in Fig. \ref{Fig1.eps}(a). 
For SHG measurements, an 800 nm ($\approx$1.55 eV), 40-fs pulse at 100 kHz from the regenerative amplifier system (RegA9040) was used as input to the optical parametric amplifier (OPA9850) to generate a signal at a wavelength of 1230 nm for a probe to induce SHG. The residual 800 nm (70-fs) pulse was used as a pump that excites the sample.
In the double-pulse TR-SHG experiment, the pump beam was split into two by a Michelson interferometer, and the time interval between the first and second pulses ($\Delta t$) was controlled by a linear stage with 10 $\mu$m resolution \cite{hase2015femtosecond}.
The ratio of the fluences of the two pump pulses was set to approximately 1:1.
In TR-SHG measurements, the pump and probe pulses were focused onto the sample normally and at a 45$^{\circ}$ incident angle, respectively, with a diameter of $\approx$60 $\mu$m and $\approx$100 $\mu$m. The pump fluence of a single pulse was varied from 1 to 6 mJ/cm$^{2}$ to prevent sample damage, while the probe fluence was fixed at 3.2 mJ/cm$^{2}$. The reflected probe (1230 nm) and SHG (615 nm) beams were directed to short-pass and band-pass filters, and only the SHG signal was delivered to fiber-coupling optics. The SHG signal was then delivered to a Si-photodiode through a multi-mode fiber (core diameter 400 $\mu$m, NA=0.39).
To obtain a time dependence of the SHG signal, the time delay ($t$) between the pump and probe pulses is modulated at 9.5 Hz by a shaker. 
All measurements were performed under ambient conditions at room temperature.

Before TR-SHG measurements, static SHG measurements were performed to clarify the SHG response from $T_{d}$-WTe$_{2}$. 
Figure \ref{Fig1.eps}(b) shows the spectrum observed for the incoming probe photon (1230 nm) and the resulting SHG signal (615 nm) emitted from the sample. Figure \ref{Fig1.eps}(c) shows the dependence of the polarization angle on the 1230 nm light pulses measured without pump light to investigate the response of SHG to crystal orientation.
SHG intensity reaches its maximum for the $b$-axis polarization and minimum for the $a$-axis polarization. For TR-SHG measurements, the polarization was set to the $b$-axis to maximize the SHG intensity.
Here, the intense probe (3.2 $\mathrm{mJ/cm^2}$) was used to obtain a sufficient signal-to-noise ratio due to the weak second harmonic effect.
It is noted that the intense probe does not influence PIPT because the fluence was selected within the region where the intensity of SHG, $I_{\mathrm{SHG}}$, was proportional to the square of the intensity of the incident probe $I_{\mathrm{probe}}$, i.e., $I_{\mathrm{SHG}} \propto I_{\mathrm{probe}}^2$ up to $\sim$3.5 mJ/cm$^{2}$ (see details in the Supplemental Material \cite{Supplemental}), as expected for a second-order nonlinear process. 
If the probe contributes to the PIPT, the SHG should decrease following the lattice symmetry changes \cite{shi2019terahertz,Zhang2019PRX}. However, this is not the case here.
Furthermore, according to the absorption coefficient obtained from our spectroscopic ellipsometry measurement, the effective photogenerated carrier density can be estimated to be $\sim$ 4.3 times larger for the pump than for the probe at the same fluence level (see the Supplemental Material \cite{Supplemental}). 
This additional measurement further supports the conclusion that the intense probe does not influence PIPT. 

\begin{figure}
   \begin{center}
	\includegraphics[width=8.5cm]{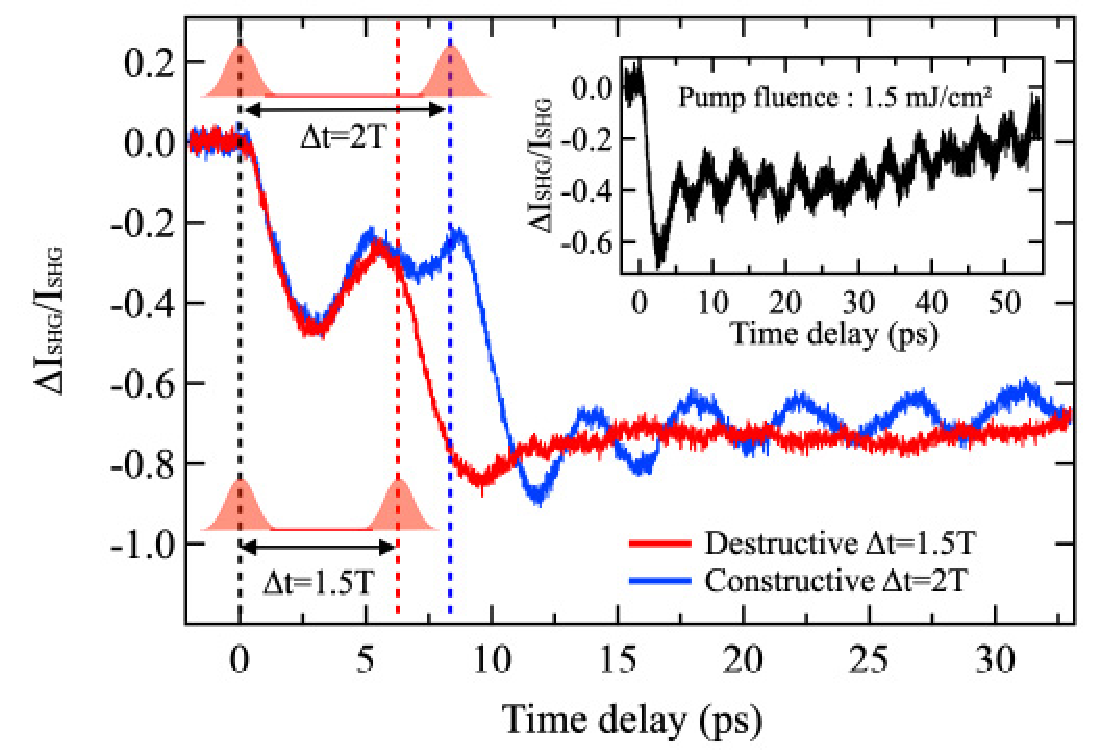}
	\caption{
    Time-domain signal of normalized change of $\Delta I_{\mathrm{SHG}}/I_{\mathrm{SHG}}$  obtained for the double pulse excitation at a total fluence of 4.6 mJ/cm$^{2}$. The blue and red lines represent the case for constructive and destructive excitation, respectively. The Gaussian-shape train indicates the arrival time of the double pulses.
    The inset shows the time-domain signal of $\Delta I_{\mathrm{SHG}}/I_{\mathrm{SHG}}$ under the single-pulse excitation at the fluence of 1.5 mJ/cm$^{2}$.
    }
	\label{Fig2.eps}
	  \end{center}
\end{figure}

To explore the contribution of electronic excitation and shear phonon oscillation in structural change, we performed coherent control of the interlayer shear phonon. 
Figure \ref{Fig2.eps} shows the time-domain signal of the SHG intensity change ($\Delta I_{\mathrm{SHG}}/I_{\mathrm{SHG}}$) obtained under double-pulse excitation together with the case of single-pulse excitation at 1.5 mJ/cm$^{2}$ (see the inset). As seen in the inset, the SHG intensity decreases just after photoexcitation ($\leq$2 ps) and then exhibits a long-lived coherent oscillation for tens of picoseconds. 
The frequency of the long-lived coherent oscillation was found to be $\approx$0.24 THz, matching the interlayer shear phonon assigned as the $A_{1}$ mode \cite{sie2019ultrafast, Ji2021ACS, aoki2022excitation}. 
The time interval between the first and second pulses $\Delta t$ is 1.5$T$ (= 6.25 ps) and 2$T$ (= 8.33 ps) of the interlayer shear phonon at the identical total pump fluence of $\approx$4.6 mJ/cm$^{2}$, where $T$ = 4.17 ps is the period of the interlayer shear phonon. 
Note that the oscillation amplitude of the higher frequency $A_{1}$ optical mode (2.4 THz) is estimated to be $\sim$10 times smaller than that of the shear mode \cite{Soranzio2022}. Therefore, the modulation of SHG intensity via the higher frequency $A_{1}$ optical mode would be too weak to be observed. 

The decrease in SHG intensity indicates that the lattice symmetry changes from $T_{d}$ toward 1$T^{\prime}$ phases \cite{sie2019ultrafast}. 
When the value of $\Delta t$ is an integral multiple of the phonon period, the oscillation amplitude of the coherent phonon generated by the first pulse is enhanced by the second pulse \cite{hase1996APL}. In contrast, when it is a half-integer multiple of the phonon period, it is canceled by the second pulse. As expected, the shear phonon was canceled when $\Delta t$=1.5$T$ and enhanced when $\Delta t$=2$T$. 
From Fig. \ref{Fig2.eps}, regardless of constructive or destructive excitation, no significant difference was observed in the decrease of the non-oscillatory component of $\Delta I_{\mathrm{SHG}}/I_{\mathrm{SHG}}$ after the arrival of the second excitation pulse (10 ps or later). 
Therefore, in the symmetry change from the $T_{d}$ toward 1$T^{\prime}$ phases under the near-infrared (NIR) photon pump (800 nm), the effect of electronic excitation is a more dominant factor than the displacement of the shear phonon. In other words, the excitation of the coherent shear phonon can be independent of the formation of the metastable phase 
after the arrival of the second pump pulse ($\geq$10 ps). 
Although decoupling of the shear phonon from the symmetry change from the $T_{d}$ toward 1$T^{\prime}$ phases was suggested by suppressing the shear phonon with lattice defects \cite{Ji2021ACS}, our results indicate more direct evidence by using coherent control of the phonon amplitude.

\begin{figure}
   \begin{center}
	\includegraphics[width=8.5cm]{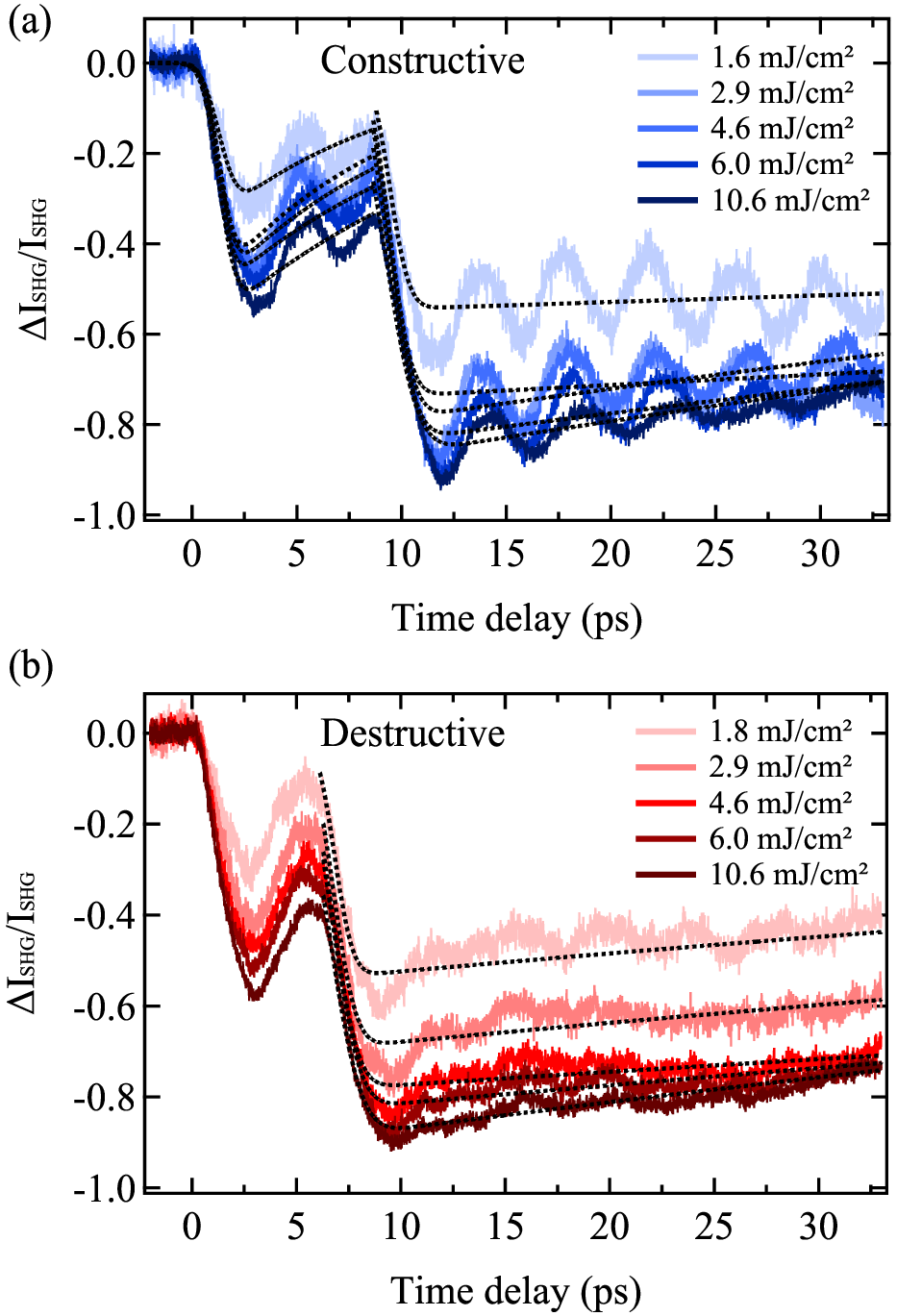}
	\caption{
	(a) Time-domain SHG intensity for constructive excitation of the shear phonon using double-pulse with $\Delta t$=2$T$ at the total fluence from 1.6 mJ/cm$^{2}$ to 10.6 mJ/cm$^{2}$. 
	(b) Time-domain SHG intensity for destructive excitation of the shear phonon with double-pulse with $\Delta t$=1.5$T$ at the total fluence from 1.8 mJ/cm$^{2}$ to 10.6 mJ/cm$^{2}$.
    The dotted lines are the fit using exponential decay functions described in the main text. 
}
	\label{Fig3.eps}
	  \end{center}
\end{figure}

To gain more insight into the effects of coherent control of the phonon amplitude on the possible lattice symmetry change from the $T_{d}$ toward 1$T^{\prime}$ phases, we present time-domain SHG signals for both constructive ($\Delta t$ = 8.33 ps) and destructive ($\Delta t$ = 6.25 ps) excitation at the total pump fluence from 1.6 (1.8) mJ/cm$^{2}$ to 10.6 mJ/cm$^{2}$ in Fig. \ref{Fig3.eps}. The amplitude of the shear phonon is enhanced by constructive excitation, as shown in Fig. \ref{Fig3.eps}(a), whereas it is canceled by destructive excitation as shown in Fig. \ref{Fig3.eps}(b). In both cases, the SHG intensity decreases as the total pump fluence increases. 

To estimate the magnitude of $\Delta I_{\mathrm{SHG}}/I_{\mathrm{SHG}}$ for single- and double-pulse excitation, the time-domain data were fit separately using an exponentially decaying function $f(t)=H(t)A\exp (-t/\tau)$, where $H(t)$ is the Heaviside function convoluted with Gaussian to account for the finite time resolution, $A$ is the magnitude, and $\tau$ is the relaxation time constant. Note that the maximum normalized change of $\Delta I_{\mathrm{SHG}}/I_{\mathrm{SHG}}$ observed at the highest total fluence of 12 mJ/cm$^{2}$ was $\approx$90\%, which is comparable to that followed by the NIR pump pulse (2.1 $\mu$m) with 10 MV/cm \cite{sie2019ultrafast}.

\begin{figure}
   \begin{center}
	\includegraphics[width=8.0cm]{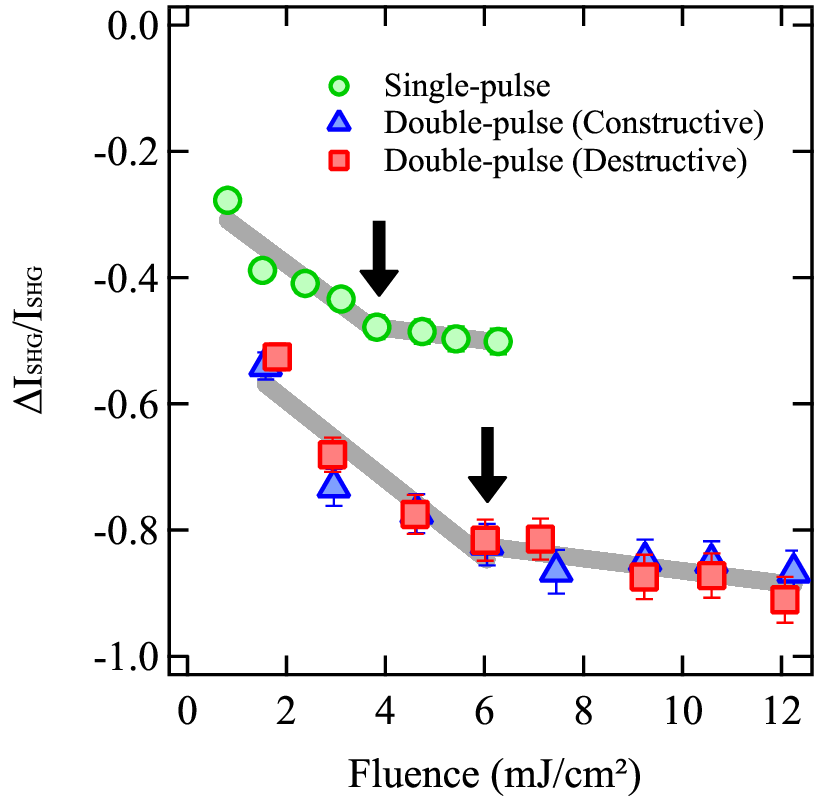}
	\caption{
	  The pump fluence dependence of the normalized SHG intensity change. The green closed circles show single-pulse excitation, while the closed blue triangles and red squares show the case for the constructive ($\Delta t$=2$T$) and destructive ($\Delta t$=1.5$T$) excitation of the shear phonon by the double-pulse, respectively. The gray thick lines represent the two different slopes obtained by the linear fits. The arrows represent the critical fluence for double-pulse excitation ($\sim$6 mJ/cm$^{2}$) and single-pulse excitation ($\sim$4 mJ/cm$^{2}$). The error bars represent fluctuations of the laser output from the amplifier ($\pm$ 4\%).
      }
	\label{Fig4.eps}
	  \end{center}
\end{figure}

Figure \ref{Fig4.eps} shows the pump fluence dependence of 
the non-oscillatory $\Delta I_{\mathrm{SHG}}/I_{\mathrm{SHG}}$ obtained from the fitting in Fig. \ref{Fig3.eps}. 
As seen in Fig. \ref{Fig4.eps}, a higher pump fluence leads to a more significant decrease in the SHG intensity for both single- and double-pulse excitation.
Moreover, even in the fluence dependence, the SHG intensity is significantly lower for double-pulse excitation than for single-pulse excitation.
This suggests that saturable absorption was suppressed by multiple-pulse excitation \cite{HaseAPL2003, SelenePRB2018, iwasaki2023electronic} and/or that the structure is more likely to change if a second pump pulse is applied after the first has already modified the structure \cite{hase2015femtosecond}.
Although there has been a Landau theory to account for the temperature dependence of the SHG intensity \cite{Harter2018PRL}, to our knowledge, an appropriate model for the fluence dependence will not currently be available under double-pulse excitation.
Thus, the fluence dependence can be adapted to two linear functions with different slopes, implying a change of the potential energy surface \cite{Ning2022PRB,KOSHIHARA20221}, e.g., a shift of the potential energy minimum toward the 1$T^{\prime}$ phase, under PIPT. In this case, the critical fluence will be higher for double-pulse excitation ($\sim$6 mJ/cm$^{2}$) than for single-pulse excitation ($\sim$4 mJ/cm$^{2}$).

To evaluate the impact of thermal effects during the phase transition from the Weyl semimetal $T_{d}$ to semimetal 1$T^{\prime}$ phases, we have calculated the electron and lattice temperatures using the two-temperature model (TTM) \cite{Allen1987PRL} under double-pulse excitation as well as incorporating possible cumulative effects \cite{weber2014heat}. 
TTM, which applies to zero-gap metallic systems, can be used to examine the effect of elevated lattice temperatures. 
The TTM results for the two typical pump fluences in our experiment (7.0 mJ/cm$^{2}$ and 9.2 mJ/cm$^{2}$) are shown in the Supplemental Material \cite{Supplemental}. 
Although the TTM analysis can only be used as a crude estimate, after the arrival of the second pulse, the lattice temperature ($T_{l}$) exceeds the transition temperature ($T_{c}$ = 565 K \cite{Tao2020PRB} or 613 K \cite{Dahal2020PRB}) at the time delay of 10 ps for 9.2 mJ/cm$^{2}$. On the other hand, $T_{l}$ is below $T_{c}$ for 7.0 mJ/cm$^{2}$. Interestingly, this critical fluence (7.0 mJ/cm$^{2}$) is close to that observed for the depletion of the SHG intensity in Fig. 4, implying that the flatter part for $\geq$ 6 mJ/cm$^{2}$ would need to be further addressed focusing on, in particular, early time dynamics (first few picoseconds) where the shear phonon displacements facilitate the symmetry switch. Note that the actual temperature increase estimated may slightly differ from the TTM results because the TTM maintains homogeneous absorption, no multi-photon absorption, both of which could occur under high-density photoexcitation, this is the case here.

Moreover, if SHG suppression is governed by cumulative heating, a single pulse with the same or higher fluence should yield similar suppression, but this is not the case here (Fig. \ref{Fig4.eps}). Thus, the cumulative heating effect would play just a minor role. 
During the first few picoseconds, the lattice absorbs energy, and the potential landscape is already distorted before the second (destructive) pulse arrives. 
The idea that destructive interference stops the shear mode tests if the larger amplitude of the shear displacement further facilitate the symmetry switch after the arrival of the second pump pulse ($\geq$10 ps). The same suppressed level of the SHG signal observed in Figs. 2, 3, and 4 for constructive and destructive excitations suggest that as the prior trajectory, just after excitation by the first pump pulse, the $T_{d}$ phase already have moved toward the 1$T^{\prime}$ phase. 
Thus, it will be required to test for the coherent control even for early time window (within a few picoseconds), although in this case we need to consider higher frequency phonons, e.g., 2.4 THz or 3.4 THz \cite{Soranzio2022}.

We argue that the lattice symmetry switch from the $T_{d}$ to the 1$T^{\prime}$ phase is driven either by interband or intraband electronic excitation. 
The former promotes electrons from bonding to anti-bonding states, increasing the charge carrier density $n_{c}$. In contrast, the latter intraband transition results in an increase in the electron temperature (or electron acceleration) without changing $n_{c}$ and can preferably be achieved by irradiating far-infrared or THz pulses \cite{sie2019ultrafast}. 
Both excitation processes can promote carrier (hole) doping and induce shear sliding of atomic layers toward the 1$T^{\prime}$ phase \cite{Xiao2020Berry}. 
Under NIR (800 nm) conditions, the photoexcited carriers in WTe$_{2}$ are generated by interband transitions along $\Gamma - X$ and $\Gamma - Y$ directions in the electronic band structure \cite{Augustin2000PRB}. 
In fact, we measured the imaginary part ($\kappa$) of the complex refractive index of WTe$_{2}$ by spectroscopic ellipsometry, and found that the 800 nm ($\approx$ 1.55 eV) pump light used in the present study was resonant with $E_{02} = $ 1.56 eV (see details in the Supplemental Material \cite{Supplemental}), which can generate the photoexcited carrier via the interband transition. Thus, the increase in charge carrier density, e.g., $n_{c}$ $\sim$ 1.26 $\times$ 10$^{21}$/cm$^{3}$ (at 1.6 mJ/cm$^{2}$) due to interband transitions plays a dominant role in the symmetry switch.

This study demonstrates that coherent control of the shear phonon amplitude can suppress its effect, isolating the impact of electronic excitation. We find that inversion-symmetry changes in the crystal structure mainly result from electronic excitation, i.e., incoherent PIPT \cite{KOSHIHARA20221}. 
In classical (incoherent) PIPT, a new phase emerges due to energy relaxation of electronic excited states, which refers to electron-phonon scattering process (phonon emission). The time constant for this incoherent emission will be on order of the phonon period, i.e., $\sim$4 ps for the shear mode, which is longer than the initial suppression of SHG signal ($\sim$1-2 ps). Thus, the incoherent PIPT will be promoted by incoherent emission of other high-frequency phonons. 
Although previous attempts have used coherent shear phonons to induce structural transitions, we successfully decouple shear phonon motion from the $T_{d}$-to-1$T^{\prime}$ phase switch in WTe$_{2}$ through destructive excitation after the arrival of the second pump pulse ($\geq$10 ps).
Since the TR-SHG technique is not a method that directly measures lattice displacement, in particular, atomic motions during this early time window (a few picoseconds), it will be important to examine the present results using direct methods such as ultrafast electron diffraction \cite{Zong2021PRL,sie2019ultrafast} and time-resolved x-ray diffraction \cite{Lindenberg2000PRL,Tinten2003Bi}, which can measure the lattice displacement associated with the shear phonon.
Even though, our findings pave the way for the investigation of phase transitions in other 2D materials, such as MoTe$_{2}$, using constructive and destructive excitation of coherent shear phonons under electronic excitation.

In conclusion, we demonstrate coherent control of coherent shear phonons in the time-domain using pairs of femtosecond laser pulses to examine the role of interlayer shear phonons on the lattice symmetry switching in solids. Combining double-pulse excitation and time-resolved SHG techniques in WTe$_{2}$, we reveal that the $T_{d}$-to-1$T^{\prime}$ symmetry switch is independent of the coherent shear phonon motion after the arrival of the second pump pulse. 
Furthermore, on the basis of spectroscopic ellipsometry measurements, we establish that electronic excitation, particularly interband transitions, plays a central role in the phase transition, reinforcing its classification as incoherent PIPT. These results provide a framework for understanding the role of shear phonons in structural changes in 2D materials through coherent control of interlayer phonons.

\section*{Acknowledgement}
This work was supported by JSPS KAKENHI (Grant Numbers. 25KJ0687, 25K17332, 23K22422 and 22KJ0352) and CREST, JST (Grant Number JPMJCR1875).
T.F. acknowledges the support of the Sasakawa Scientific Research Grant from the Japan Science Society.
Y.M. acknowledges the support from JST SPRING, Japan Grant Number JPMJSP2124.
M.H. acknowledges the support from the DGIST R\&D program (24-KUJoint-06).
We thank Sathvik A. Iyengar and Gyan Prakash for their critical reading of the manuscript and insightful feedback, and Masaki Hada and the Organization for Open Facility Initiatives, University of Tsukuba, for the spectroscopic ellipsometry measurements.

\textit{Note added.} -- After this paper was submitted, we became aware of related results by Horstmann \textit{et al}. \cite{horstmann2025} on precise coherent phonon control avoiding amplitude saturation.

\bibliography{WTe2_SHG}
\end{document}


\preprint{}

\title{\textbf{Supplemental Material for\\
"Role of interlayer shear phonons on lattice symmetry switching in the transition metal dichalcogenide WTe$_{2}$"} 
}%
\author{Mizuki Akei}
\email{s2420263@u.tsukuba.ac.jp}
\affiliation{Department of Applied Physics, Graduate school of Pure and Applied Sciences, University of Tsukuba, 1-1-1 Tennodai, Tsukuba 305-8573, Japan}
\author{Takumi Fukuda}
\email{takumi.fukuda@oist.jp}
\affiliation{Femtosecond Spectroscopy Unit, Okinawa Institute of Science and Technology Graduate University, 1919-1 Tancha, Onna, Okinawa, Japan}
\author{Yu Mizukoshi}
\affiliation{Department of Applied Physics, Graduate school of Pure and Applied Sciences, University of Tsukuba, 1-1-1 Tennodai, Tsukuba 305-8573, Japan}
\author{Kazuhiro Kikuchi}
\affiliation{Department of Applied Physics, Graduate school of Pure and Applied Sciences, University of Tsukuba, 1-1-1 Tennodai, Tsukuba 305-8573, Japan}
\author{Muneaki Hase}
\email{mhase@bk.tsukuba.ac.jp}
\affiliation{Department of Applied Physics, Graduate school of Pure and Applied Sciences, University of Tsukuba, 1-1-1 Tennodai, Tsukuba 305-8573, Japan}


\maketitle

\newpage

\section*{I. The influence of intense probe on PIPT\protect}

In the present study, an intense probe (3.2 mJ/cm$^{2}$) is needed to obtain sufficient signal-to-noise ratio because of the weak second harmonic effect. Although the intense probe pulse was used, we show in the following the measurements were performed under the condition that the intense probe did not influence the photo-induced phase transition (PIPT). First, as shown in Fig. \ref{FigS1.eps} below, the probe fluence was selected within the region where the SHG intensity follows a quadratic dependence. We should observe a decrease of the SHG signal if the PIPT could occur and the lattice symmetry changes at 3.2 mJ/cm$^{2}$, this is not the case here.

\begin{figure}[h]
\makeatletter
\renewcommand{\thefigure}{%
\thesection S\arabic{figure}}
\@addtoreset{figure}{section}
\makeatother
	\includegraphics[width=8.6cm]{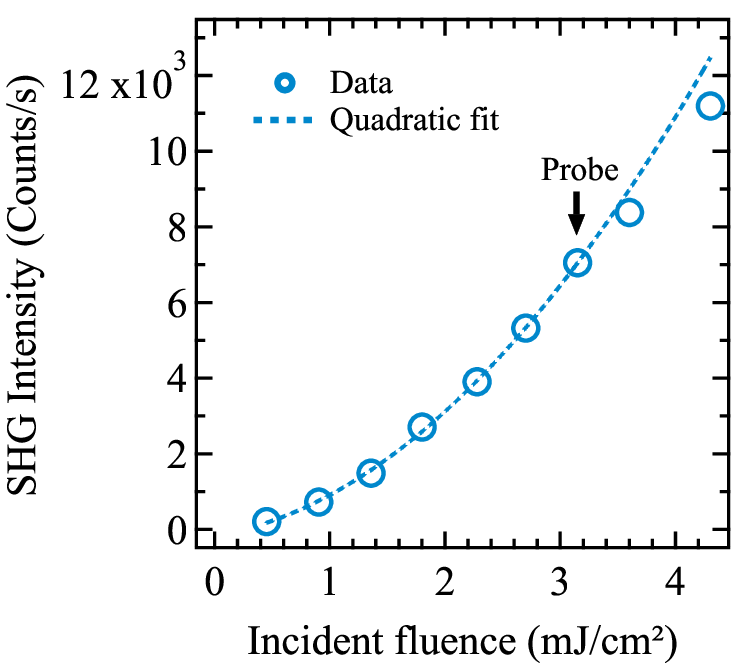}
	\caption{
	The number of SHG photons per second as a function of the incident fluence in WTe$_{2}$. The arrow shows the probe fluence (3.2 mJ/cm$^{2}$) used in the present study.}
	\label{FigS1.eps}
\end{figure}

\begin{figure}[h]
\makeatletter
\renewcommand{\thefigure}{%
\thesection S\arabic{figure}}
\@addtoreset{figure}{section}
\makeatother
	\includegraphics[width=12.6cm]{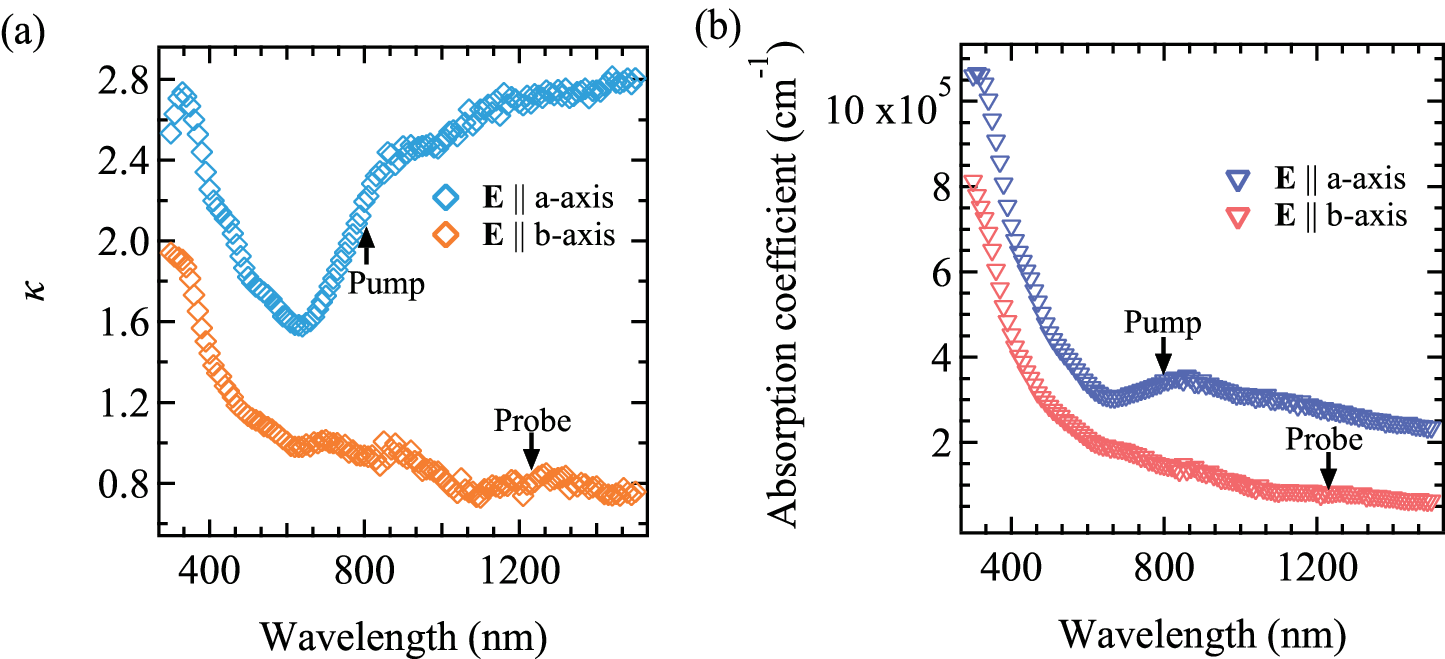}
	\caption{
	(a) The imaginary part of the complex refractive index in the crystal axes of a- and b-directions obtained in the WTe$_{2}$ sample used. (b) The absorption coefficient $\alpha$ in the crystal axes of a- and b-directions. The arrows indicate the wavelength for the pump and probe pulses.}
	\label{FigS2.eps}
\end{figure}

Secondly, we have investigated the anisotropic optical properties of WTe$_{2}$ using spectroscopic ellipsometry. In the experiment, the pump (800 nm) polarization was parallel to the a-axis, while that of the probe (1230 nm) was parallel to the b-axis, resulting in different absorption characteristics. The imaginary ($\kappa$) part of the complex refractive index and the absorption coefficient ($\alpha$) obtained by spectroscopic ellipsometry measurement are shown in Fig. \ref{FigS2.eps}. We found the value of $\alpha$ = 3.45×10$^{5}$ cm$^{-1}$ at 800 nm, whereas $\alpha$ = 0.81×10$^{5}$ cm$^{-1}$ at 1230 nm. Therefore, the absorption of pump light is $\approx$4.3 times larger than that of the probe. This means that the photogenerated carrier density is $\approx$4.3 times larger for the pump than for the probe at the same fluence level. In the present case, the photogenerated carrier density is estimated to be 1.26×10$^{21}$ cm$^{-3}$ for the lowest pump (1.6 mJ/cm$^{2}$) and to be 1.06×10$^{21}$ cm$^{-3}$ for the probe (3.2 mJ/cm$^{2}$). Thus, the photogenerated carrier density is smaller for the probe even when the lowest pump fluence was irradiated to the sample.

\section*{II. The Two-temperature model and cumulative effect\protect}
To consider the thermal effects of laser irradiation, the lattice temperature increase was estimated using the two-temperature model (TTM)\cite{allen1987theory,mondal2018coherent}.
The TTM is given by the set of two coupled heat equations,
\begin{equation}
\label{TTM_1}
C_e(T_e)\frac{\partial T_e}{\partial t}=-G(T_e-T_l)+S(t),
\end{equation}
\begin{align}
C_l\frac{\partial T_l}{\partial t}=G(T_e-T_l),
\end{align}
where $T_e$ and $T_l$ are the temperatures of the electrons and the lattice, respectively. $C_e(=\gamma T_e)$ and $C_l$ are the heat capacity of the electron and lattice, respectively. $\gamma$ is the Sommerfeld constant. We calculated the value of $\gamma$ using $\gamma=\pi^2k_B^2n_e/(2E_F)$, where $k_B$ is the Boltzmann constant, $n_e$ is the carrier density, and $E_F$ is the Fermi energy. $G$ is the $e-ph$ coupling constant. From the literature value of constants, $C_L=77.53$ JK$^{-1}$mol$^{-1}$ (Ref. \cite{callanan1992thermodynamic}) and $G\sim6.79\times10^{15}$ Wm$^{-3}$K$^{-1}$ (Ref. \cite{dai2015ultrafast}). The final term $S(t)$ in Eq. (\ref{TTM_1}) is the source term of the laser pulse that heats the sample. 
The double-pulse can be written as the combination of two Gaussian functions \cite{mondal2018coherent,yu2023anomalous,yu2023all},

\begin{align}
S(z,t)=\sqrt{\frac{4\ln(2)}{\pi}}\frac{1-R}{\delta t_p}e^{-\frac{z}{\delta}}\Big\{F_1e^{-4\ln(2)\big(\frac{t}{t_p}\big)^2}+F_2e^{-4\ln(2)\big(\frac{t-\Delta t}{t_p}\big)^2}\Big\},
\end{align}
where $R$ ($\sim0.43)$ is the optical reflectivity and $\delta$ ($\sim30$ nm) is the optical penetration depth obtained by the spectroscopic ellipsometry measurements. $F_1$ is the first incident pump fluence, $F_2$ is the second one, $t_p$ ($\sim 70$ fs) is the laser pulse width and $\Delta t$ (=6.25 ps) is the pulse interval for the case of destructive excitation.

\begin{figure}[h]
\makeatletter
\renewcommand{\thefigure}{%
\thesection S\arabic{figure}}
\@addtoreset{figure}{section}
\makeatother
	\includegraphics[width=12.6cm]{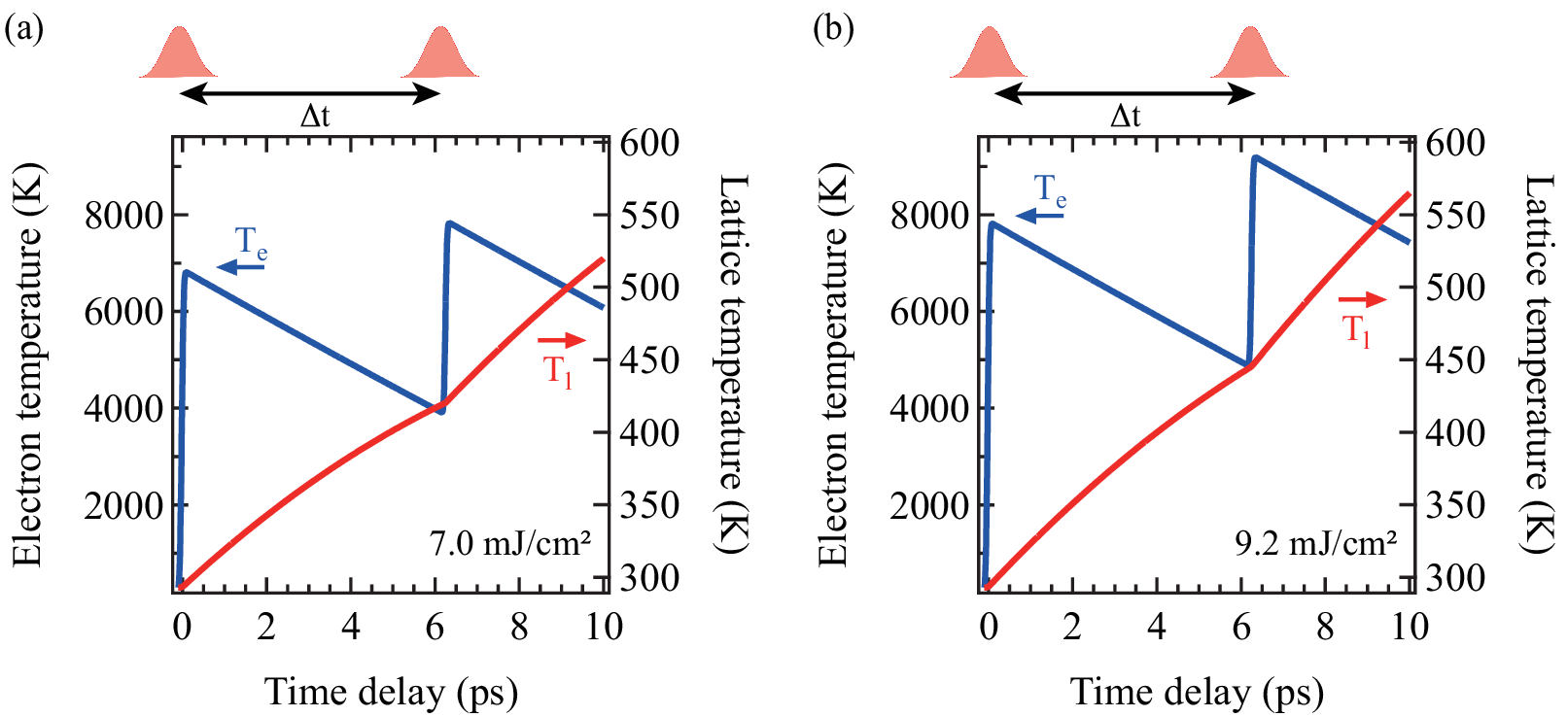}
	\caption{
	(a) The electron and lattice temperatures for double-pulse excitation at a total fluence of 7.0 mJ/cm$^{2}$ as estimated by the TTM model. The Gaussian-shape train indicates the arrival time of the double pulses with the time interval of $\Delta t$. (b) The same as (a) but for a total fluence of 9.2 mJ/cm$^{2}$.
    }
	\label{FigS3.eps}
\end{figure}

The TTM results for the two typical pump fluences in our experiment (7.0 mJ/cm$^{2}$ and 9.2 mJ/cm$^{2}$) are shown in Fig. \ref{FigS3.eps}. Although the TTM analysis can only be used as a crude estimate, after the arrival of the second pulse, the lattice temperature $T_l$ exceeds the transition temperature ($T_c$ = 565 K \cite{Tao2020PRB} or 613 K \cite{Dahal2020PRB}) at a time delay of 10 ps for 9.2 mJ/cm$^{2}$ (Fig. \ref{FigS3.eps}(b)). On the other hand, $T_l$ is below $T_c$ for 7.0 mJ/cm$^{2}$ (Fig. \ref{FigS3.eps}(a)). Interestingly, this critical fluence (7.0 mJ/cm$^{2}$) is close to that observed for the depletion of the SHG intensity in Fig. 4, implying that the flatter part for $\geq $ 6 mJ/cm$^{2}$ would need to be further addressed focusing on, in particular, early time dynamics (first several picoseconds) where shear phonon displacements facilitate the symmetry switch.

Regarding the possible cumulative effect, we have calculated it using the existing heat accumulation model \cite{weber2014heat}:
\begin{align}
\Delta{T_{Sum, 3D}(t)}=\frac{Q_{3D}}{\rho{C_L}\sqrt{(4\pi\kappa)^{3}}}\sum_{N=1}^{N_p}\frac{\Theta\Big(t-\frac{N-1}{f_L}\Big)}{\sqrt{\Big(t-\frac{N-1}{f_L}\Big)^{3}}}e^{-\frac{1}{\big(t-\frac{N-1}{f_L}\big)}\frac{{r^2_{3D}}}{4\kappa}},
\end{align}

where $Q_{3D}$ is the energy that is released in an infinitely short time at $t$ = 0, $\rho$ is the mass density of the solid, $C_L$ is the specific heat capacity, $\kappa=\lambda/(\rho C_L)$ with $\lambda$ being the thermal conductivity, $f_L$ is the repetition rate, $\Theta$ represents the Heaviside function, and $r_{3D}=\sqrt{x^{2}+y^{2}+z^{2}}$ is the distance from the surface. 
Under irradiation with the 7 mJ/cm$^{2}$ (in total) pulse fluence, the cumulative effect can be calculated to $\sim$ 15 K for 50 pulses at 100 kHz as demonstrated in Fig. S4.
\begin{figure}[h]
\makeatletter
\renewcommand{\thefigure}{%
\thesection S\arabic{figure}}
\@addtoreset{figure}{section}
\makeatother
	\includegraphics[width=8.6cm]{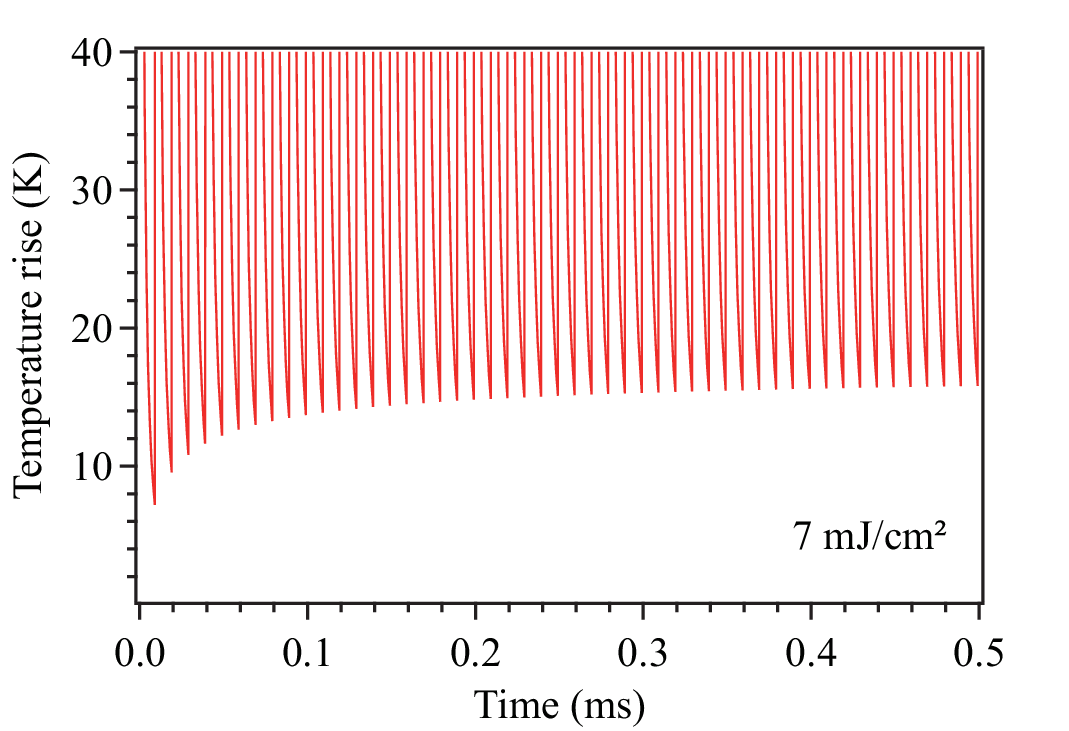}
	\caption{The estimation of the cumulative effect under irradiation by 50$\times$7 mJ/cm$^{2}$ pulses at the sample surface ($r_{3D}$=0). We used the thermal conductivity of $\lambda$=15 W m$^{-1}$ K$^{-1}$ (Ref. \cite{chen2019plane}), and the density of $\rho$=9430 kg m$^{-3}$, the lattice specific heat of $C_L$=176.6 J kg$^{-1}$ K$^{-1}$ (Ref. \cite{callanan1992thermodynamic}).}
	\label{FigS4.eps}
\end{figure}

\begin{figure}[h]
\makeatletter
\renewcommand{\thefigure}{%
\thesection S\arabic{figure}}
\@addtoreset{figure}{section}
\makeatother
	\includegraphics[width=7cm]{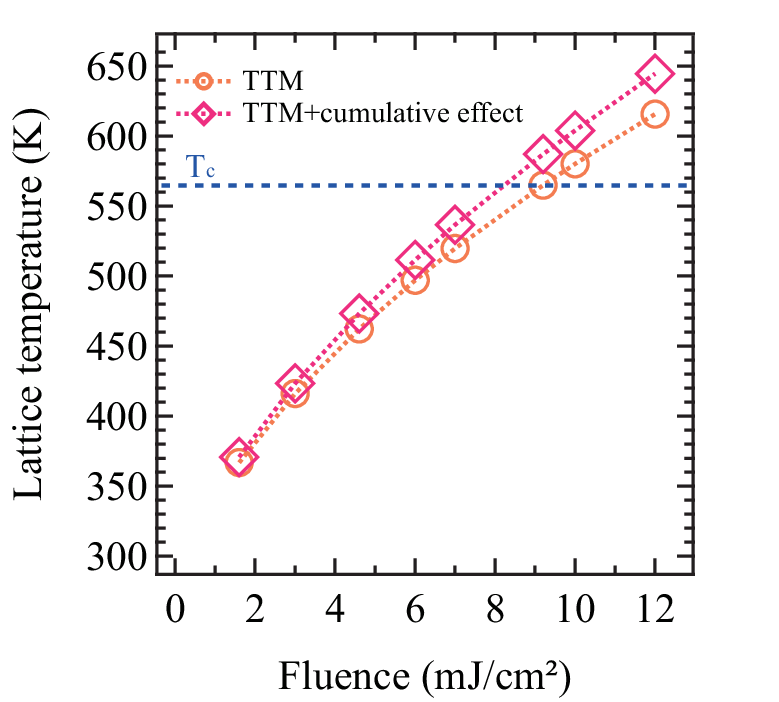}
	\caption{The estimation of the lattice temperature based on the combination of the TTM results (Fig. \ref{FigS3.eps}) and heat cumulative effect (Fig. \ref{FigS4.eps}). The value of lattice temperature increase due to the cumulative effects is selected at 3 ms (data not shown, but it is obtained by the extended data from Fig. \ref{FigS4.eps}.) The horizontal dashed line indicates the lower bound of $T_c$=565 K.}
	\label{FigS5.eps}
\end{figure}

Then we could incorporate possible cumulative effects into the TTM, as demonstrated in Fig. \ref{FigS5.eps}. Even when cumulative effects are included, the lattice temperature $T_l$ is below $T_c$ (565 K or 613 K) for 7.0 mJ/cm$^{2}$. Thus, we can conclude that the lattice symmetry switching from the Weyl semimetallic T$_d$ to the semimetallic 1T' phases is not promoted by the lattice temperature rise at the total pulse fluence of $\leq$ 7 mJ/cm$^{2}$.

\section*{III.  Imaginary part of the complex refractive index and fit by the Lorentz model\protect}

\begin{figure}[h]
\makeatletter
\renewcommand{\thefigure}{%
\thesection S\arabic{figure}}
\@addtoreset{figure}{section}
\makeatother
	\includegraphics[width=7.6cm]{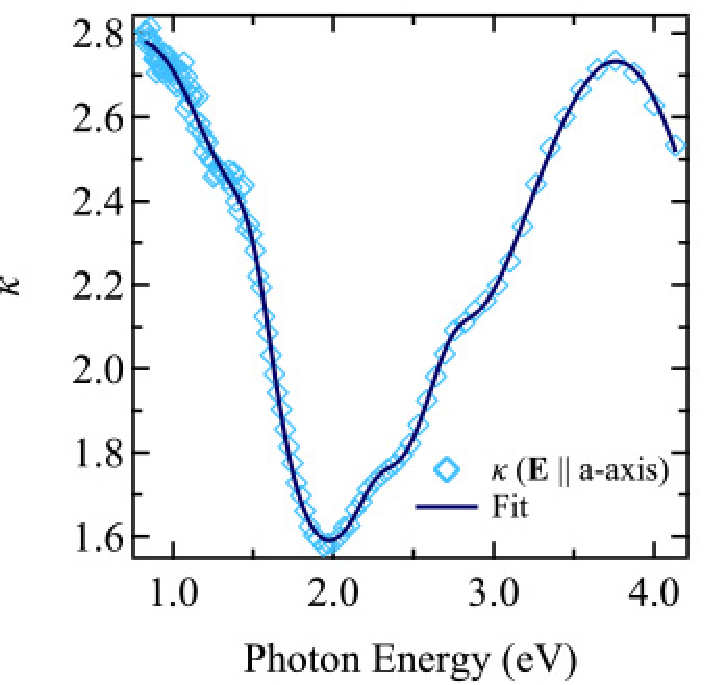}
	\caption{The imaginary part of the complex refractive index of the WTe$_{2}$ sample measured by spectroscopic ellipsometry together with the fit by the Lorentz model.}
	\label{FigS6.eps}
\end{figure}

To clarify the excitation mechanism leading to phase switching, we performed spectroscopic ellipsometry measurements to obtain the absorption spectrum. The imaginary part measured ($\kappa$) of the complex refractive index of $T_{d}$-WTe$_{2}$ is shown in Fig. \ref{FigS6.eps}, which is roughly consistent with a previous study \cite{buchkov2021anisotropic}. The experimental data can be well fitted by the Lorentz model as follows:
\begin{align}
\varepsilon(E)=\varepsilon_b+\sum_{j}\frac{C_j}{E^2-E_{0j}^2-iE\gamma_j}.
\end{align}
Here, $\varepsilon(E)$ is a complex dielectric function, $E$ is the photon energy in eV, $C_j$ is the amplitude of the $j$-th resonance, $E_{0j}$ is the $j$-th resonant energy, $\gamma_j$ is the $j$-th linewidth, and $\varepsilon_b$ is the high frequency limit of the dielectric function. The parameters of the Lorentz model are shown in Table 1. According to the Lorentz fit, a resonance peak is found to exist at 1.56 eV, corresponding to the interband transition.

\begin{table}[b]
\caption{
The coefficients obtained from the fitting of the imaginary part using the Lorentz model in Fig. \ref{FigS6.eps}. $\varepsilon_b =$ 16.9 was obtained.}
\begin{ruledtabular}
\begin{tabular}{cccccc}
 $j$ & 1 & 2 & 3 & 4 & 5 \\
\hline
$C_j$ (eV$^2$) & 17.7 & 2.64 & 1.13 & 2.47 &  209 \\
        $E_{0j}$ (eV) & 1.32 & 1.56 & 2.31 & 2.79 & 4.23 \\
        $\gamma_{0j}$ (eV) & 1.45 & 0.46 & 0.43 & 0.50 & 2.14 \\
\end{tabular}
\end{ruledtabular}
\end{table}

\clearpage

\nocite{*}

\bibliography{RefSM0918}